\documentclass[vecphys]{svmult}

\usepackage{makeidx}         
\usepackage{graphicx}        
\usepackage{multicol}        
\usepackage[bottom]{footmisc}

\makeindex             

\begin{document}

\title*{Near UV properties of Early--Type Galaxies at $z\sim1$}
\author{Sperello di Serego Alighieri\inst{1}\and
Sandro Bressan\inst{2}}
\institute{INAF -- Osservatorio Astrofisico di Arcetri, Largo E. Fermi 5,
50125 Firenze, Italy
\texttt{sperello@arcetri.astro.it}
\and INAF -- Osservatorio Astronomico di Padova, Vicolo dell'Osservatorio
5, 35100 Padova, Italy}
%
%
\maketitle

\begin{abstract}
We have used spectral fits to SSP--based atmosphere models to derive an
estimate of the average stellar age for an almost complete sample of 15
Early-Type Galaxies (ETG) at $0.88<z<1.3$. The results are in only partial
agreement with the age estimates previously obtained for the same objects
from an analysis of the ${\cal M}/L_B$ ratio, derived from the Fundamental
Plane (FP) parameters. In particular spectral fits seem to underestimate the
age of the most luminous ETG, and therefore do not reproduce the
downsizing effect, which is clear for the FP ages. We also analyse the
relationship between the spectral--fit ages and various near--UV spectral
indices.
\end{abstract}

\section{Introduction}
\label{Intro}
The determination of the age of Early--Type Galaxies (ETG) as a function
of galaxy mass and environment is crucial for the understanding of galaxy
formation and evolution and provides the means of
discriminating between the currently competing models of galaxy evolution
\cite{ren06}.
Using the dynamical, morphologic and photometric parameters which are used
for the Fundamental Plane (FP) \cite{djo87}, di Serego Alighieri, Lanzoni
\& J\o rgensen \cite{dis06} have made an estimate of the age of an almost
complete sample of 15 ETG at $0.88<z<1.3$ selected from the K20 survey
\cite{dis05}. This age estimate was based on an evaluation of the virial
mass ${\cal M}$ and rested on interpreting the differences in ${\cal M}/L_B$
as age differences, using single stellar population synthesis models.
Therefore luminosity--weighted average stellar ages were obtained.

The result of this analysis is that the age of ETG increases with the
galaxy mass in all environments (the so--called {\it downsizing} \cite{cow96})
and that cluster galaxies appear to have the same age, within 5\%, as field
galaxies at any given galaxy mass. The second results is important, since
it is contrary to the predictions of the most recent incarnation of the
hierarchical models of galaxy formation and evolution \cite{del06}, and it
is somewhat controversial \cite{tho05}. In order to check its validity, we
have looked for an independent estimate of galaxy ages and present here
the preliminary results of our study. 

\section{Galaxy ages from spectral fits}
\label{fits}
\begin{figure}
\centering
\includegraphics[height=11cm]{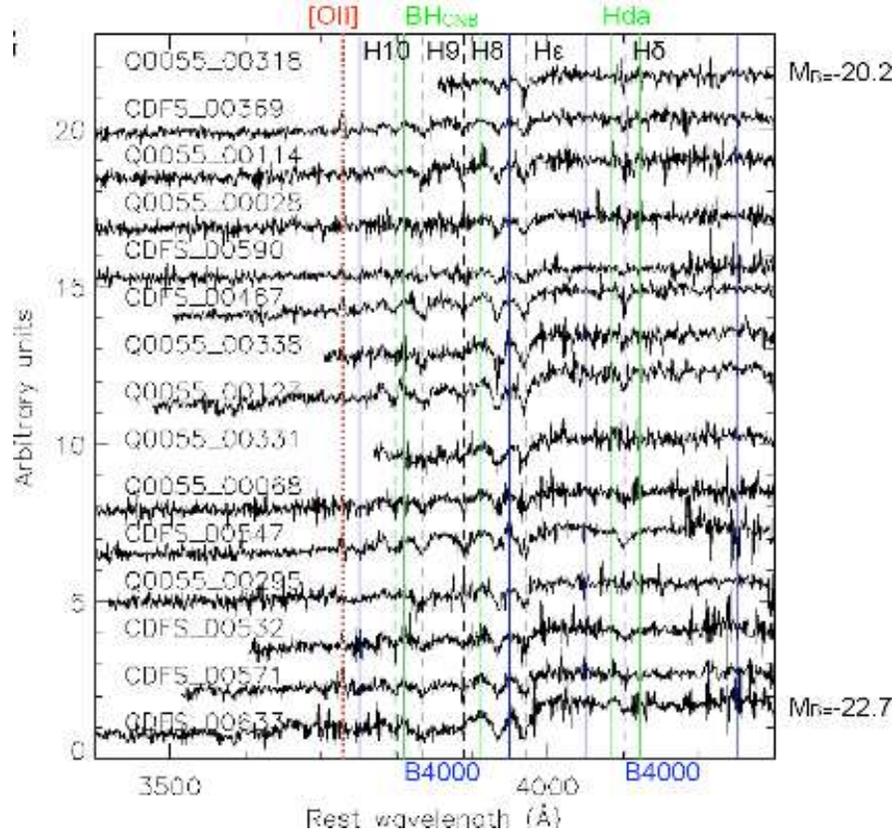}
%
%
\caption{The intermediate resolution spectra of the 15 ETG at $z\sim1$
from the K20 survey. The spectra are ordered with galaxy luminosity
decreasing upward. A number of important spectral features are marked,
with their wavelength range. The ETG, for which the spectral--fit age is
lower than the FP age, are marked with arrows.
}
\label{fig:1}       
\end{figure}

Figure 1 shows the rest--frame blue and near UV part of the spectra of our 
sample of 15 ETG at $0.88<z<1.3$, which have been used to derive the velocity 
dispersion \cite{dis05}. These spectra have a resolution of
$\Delta\lambda /\lambda = 1400$ and have a S/N ratio variable between 20
and 40. They are therefore suitable for stellar population analysis. We
have fitted them with high resolution SSP--based atmosphere models by
Bertone et al. (in prep.) in order to derive an estimate of the average
SSP age. To break the age--metallicity degeneracy, we have adopted the
metallicity derived from the measured velocity dispersion using an
empirical relationship \cite{ann07}.
We stress that, although both age determination methods are based on the 
same spectra, they are rather independent.

\begin{figure}
\centering
\includegraphics[height=10cm]{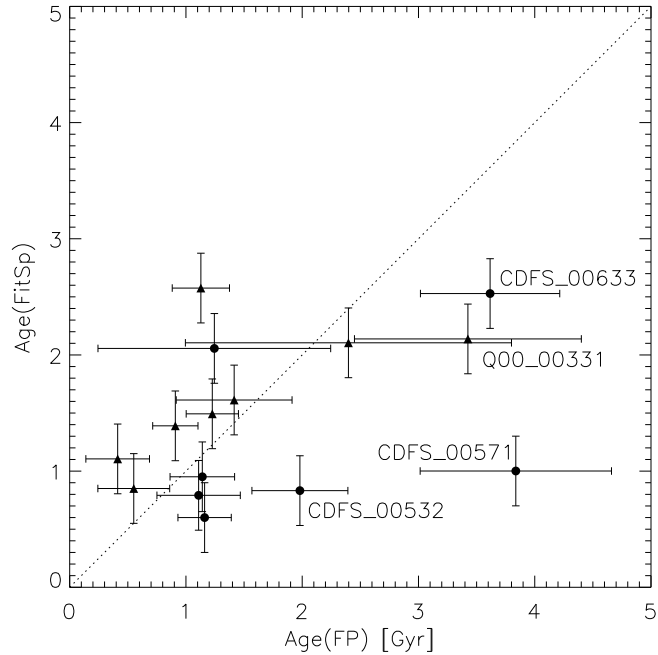}
%
%
\caption{A comparison of the ages obtained from the FP parameters with
those obtained from the spectral fits. In both cases we have used
metallicities derived from the velocity dispersion \cite{ann07}.
Some objects with discrepant ages are labeled.
}
\label{fig:2}       
\end{figure}

Figure 2 shows the comparison between the ages obtained from the FP
parameters and the analysis of the ${\cal M}/L_B$ ratio with those
obtained by fitting the spectra with SSP models. We note that the
discrepant objects, which are marked by arrows in Fig. 1, are the
most luminous ones. In fact the spectral fitting technique gives ages
between 0.5 and 2 Gyr for all the ETG, including the brightest ones,
while the techique based on the ${\cal M}/L_B$ ratio assigns larger ages
to the most luminous ETG. Although we cannot assess with certitude which
method is best, we note that, adopting the spectral--fit ages, most of
the downsizing effect would disappear (see figure 3).

\begin{figure}
\centering
\includegraphics[height=5.5cm]{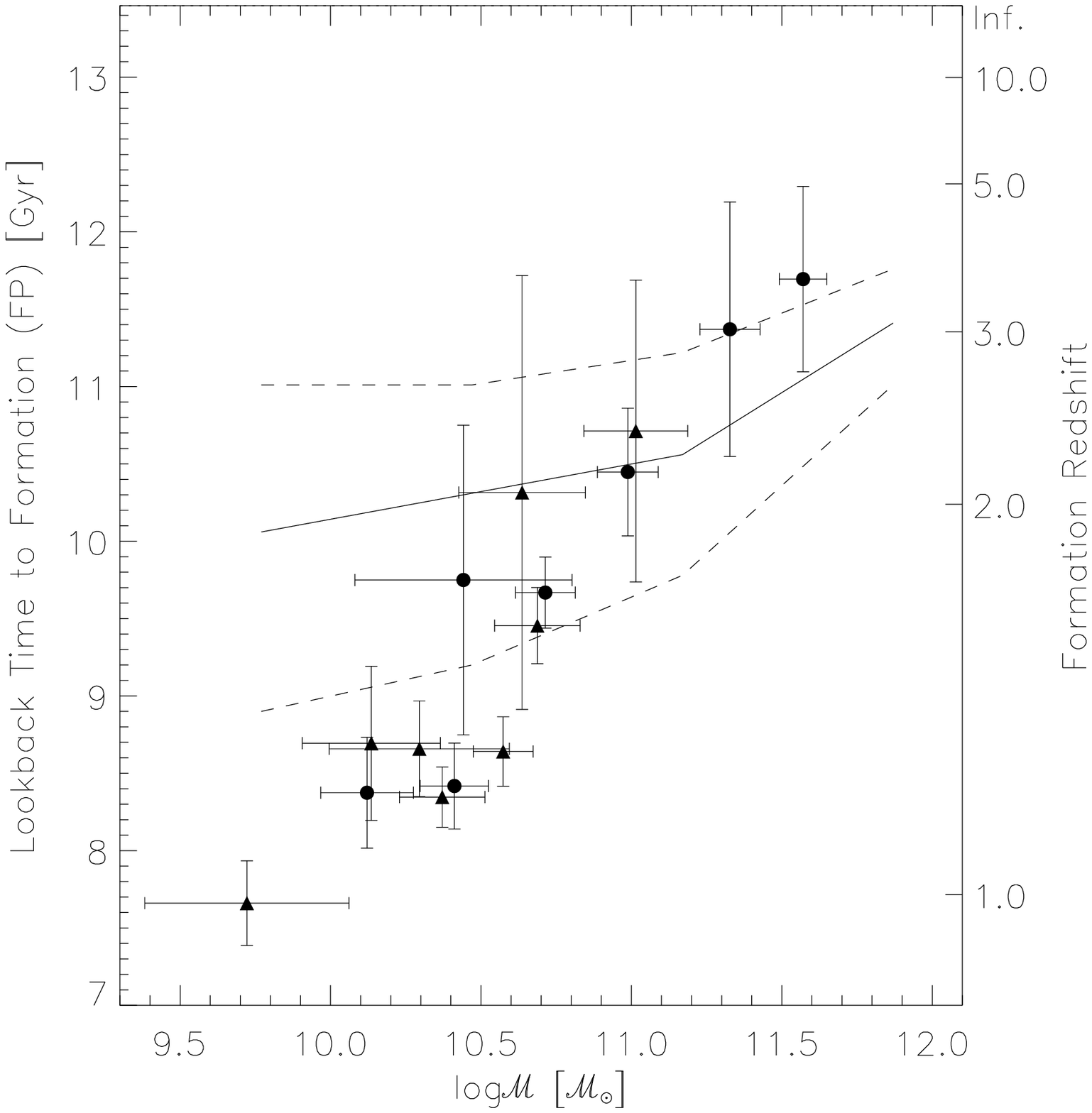}
\includegraphics[height=5.5cm]{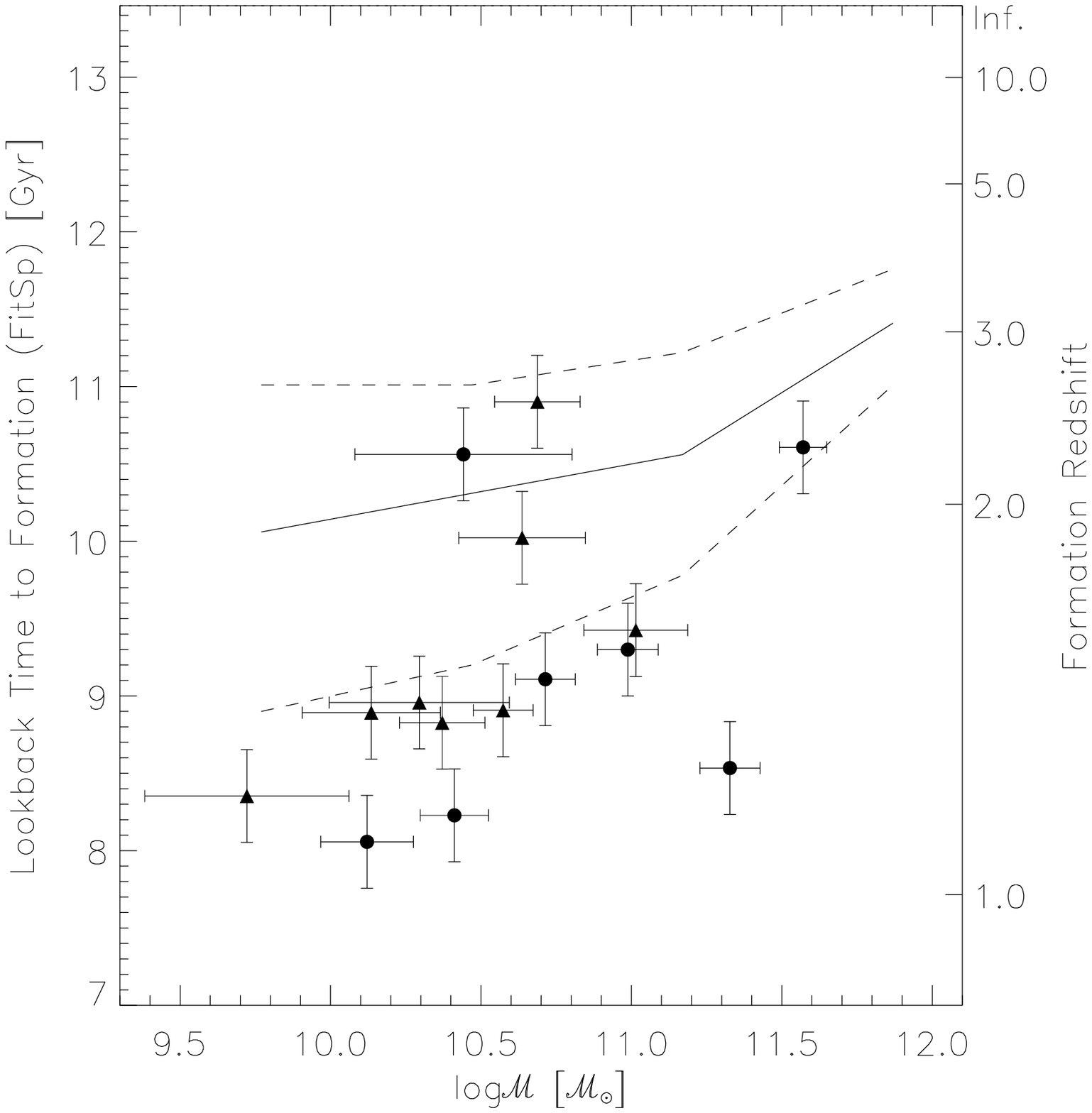}
\caption{The dependence of the lookback time to formation with the galaxy
mass: on the left the age is estimated from the FP parameters, while on
the right it is estimated from spectral fits. The downsizing effect is
clear on the left, but not on the right. The continuous line shows the
median model ages obtained from a semianalytical model of hierarchical
galaxy evolution, while the dashed lines are their upper and lower
quartiles \cite{del06}.
}
\label{fig:3}       
\end{figure}

\section{Spectral indices}
The blue/near--UV region of the spectrum contains several important
spectral indices \cite{bro86}, some of which are marked 
in Fig. 1. These are also good indicators of the stellar population 
content of each galaxy and offer therefore a complementary information 
which is potentially usefull for the age determination \cite{lon00}.
We show in Fig. 4 the dependence of these spectral indices
with the age estimated from the spectral fits.

\begin{figure}
\centering
\includegraphics[height=5.5cm]{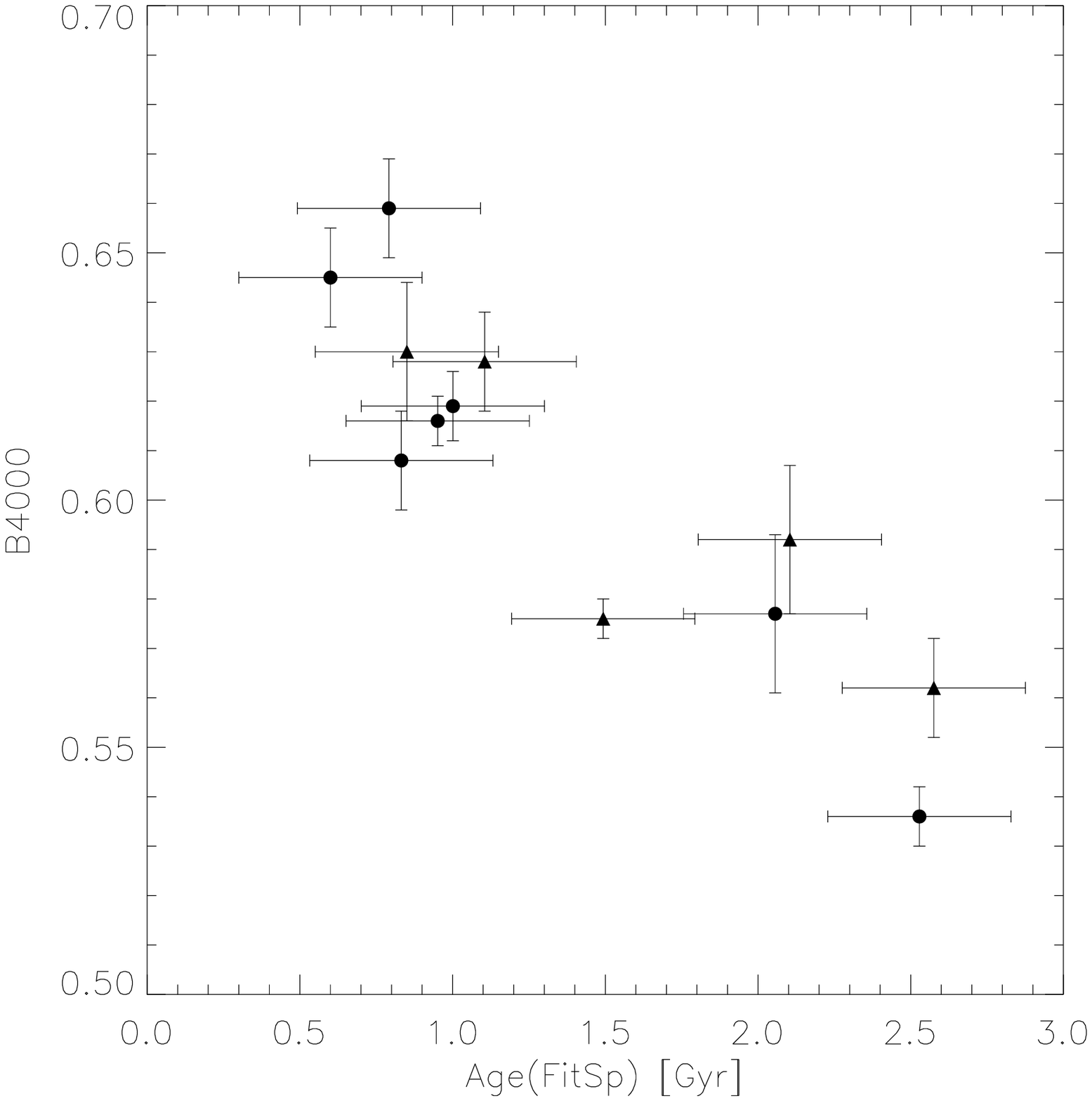}
\includegraphics[height=5.5cm]{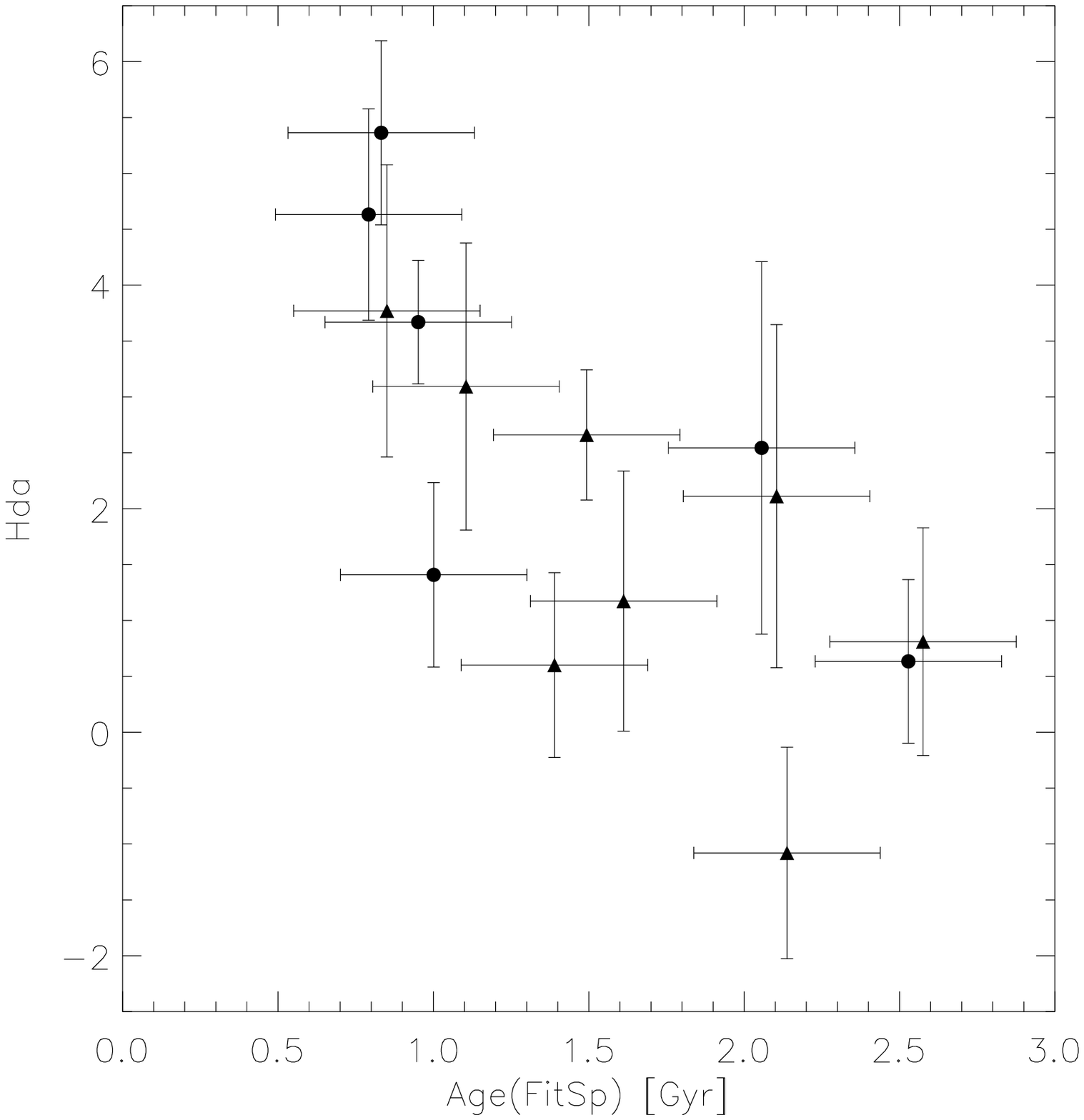}
\includegraphics[height=5.5cm]{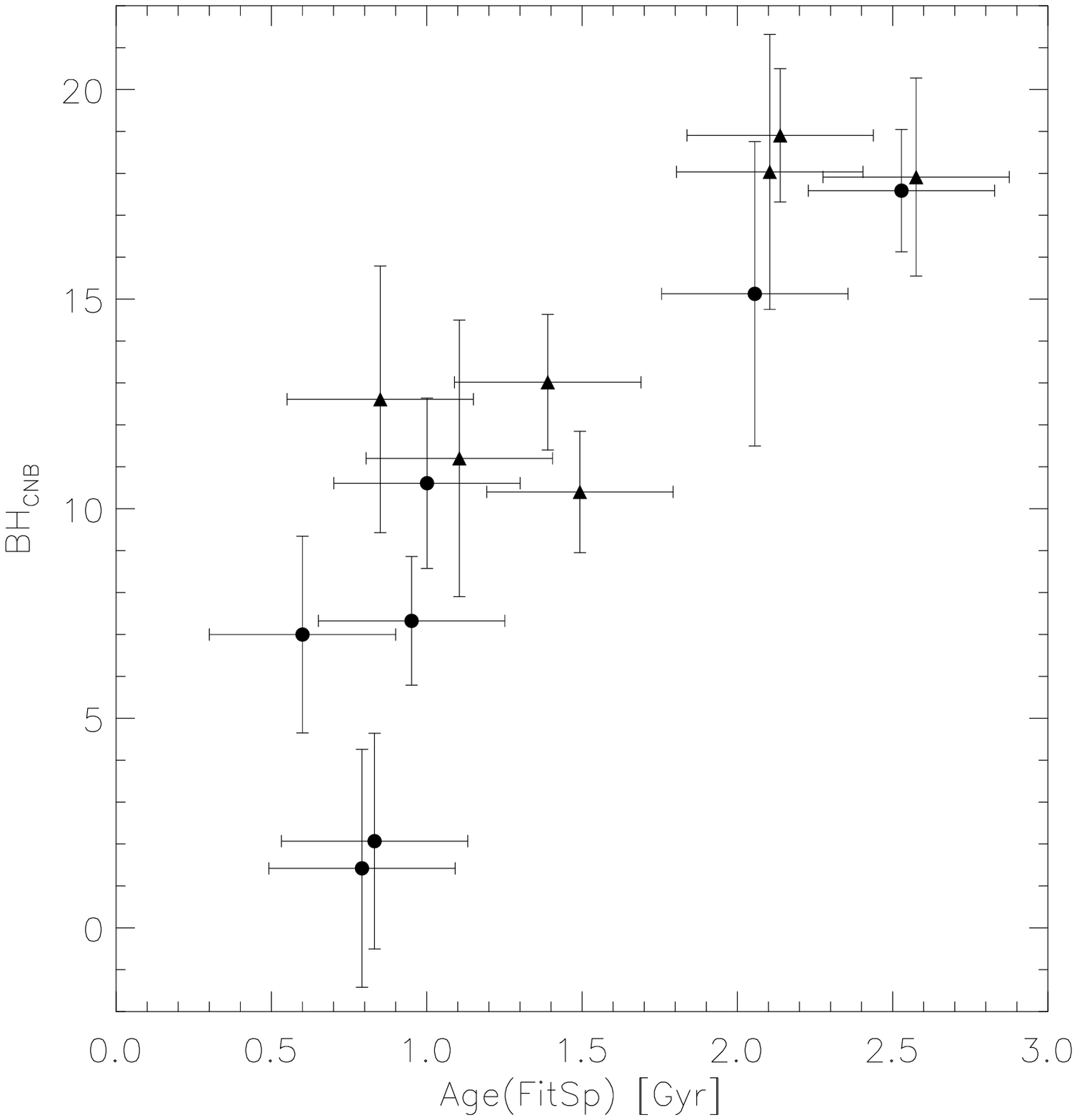}
\includegraphics[height=5.5cm]{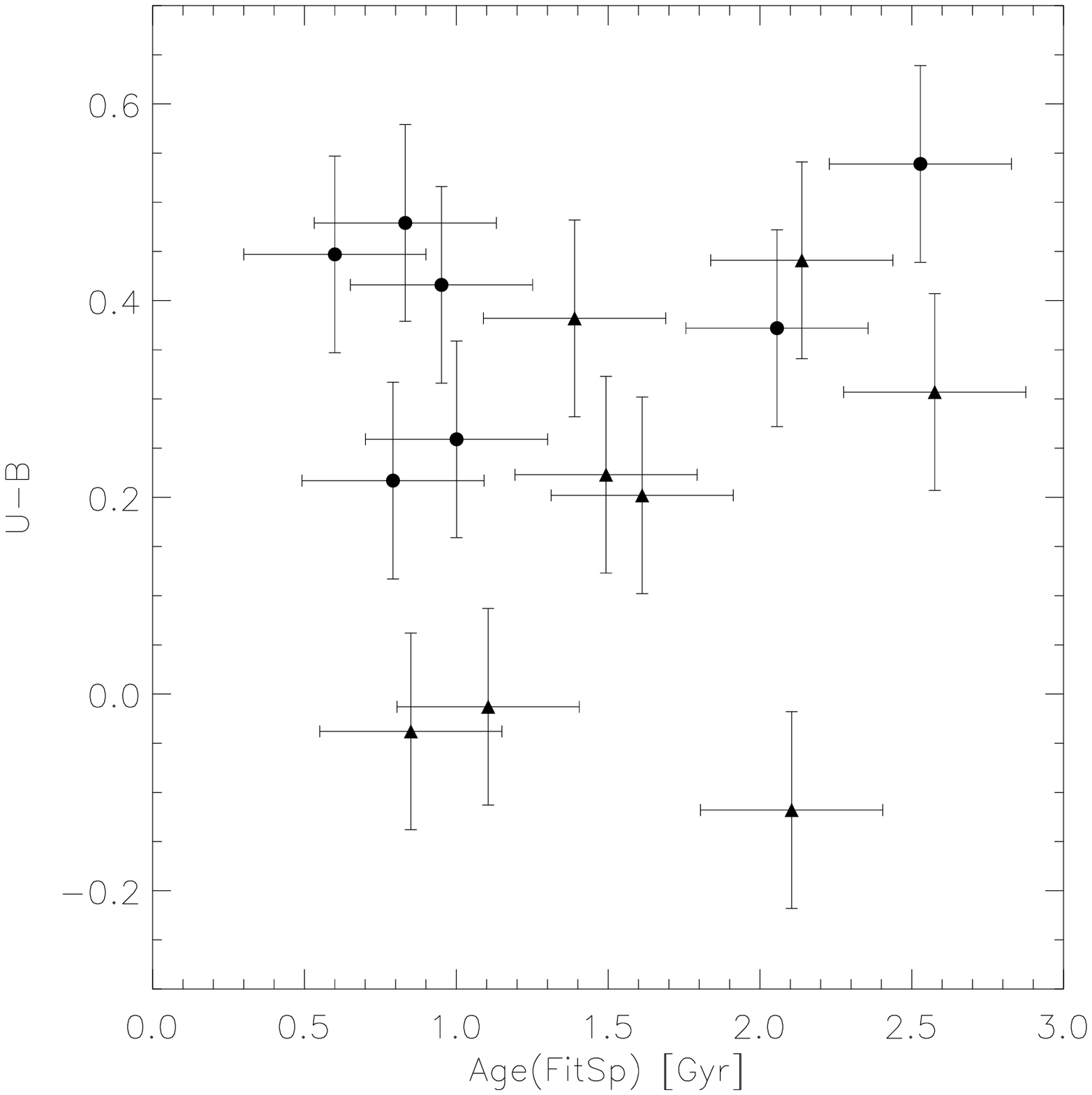}
\caption{The  dependence of some blue/near--UV spectral indices and colour
with the spectral--fit ages.
}
\label{fig:4}       
\end{figure}

It is interesting, although not completely surprising, that there is a
good correlation between ages and spectral indices. On the other hand the
U-B colour does not seem to correlate with the spectral--fit age.

\section{Conclusions}
A comparison between the ages of a sample of 15 ETG at $z\sim 1$ estimated
using spectral fits to atmosphere models with those derived from the FP
parameters shows a disagreement for the most luminous galaxies, which
appear younger with the spectral fit method, thereby destroying the 
downsizing effect with this method.
We suggest that spectral--fit ages are quite sensitive to rather small
amounts of current or recent star formation activity, which involve only a
small percentage of the stellar mass of the galaxy. This would be
consistent with the good relationship found with the near--UV spectral
indices. On the other hand FP ages seem to be more relevant for the
average stellar population of ETG at $z\sim 1$.

%
%



\printindex
\end{document}